\documentclass[aps,twocolumn,showpacs,amsfonts,amssymb,amsmath]{revtex4}  

\usepackage{graphicx}

\newcommand{\EXP}[1]{\mathrm{e}^{#1}} 

\newcommand{\DEFt}{\smash{\overset{\text{\tiny def}}{=}}}
\newcommand{\imat}{{\mathrm{i}}} 
\newcommand{\tr}{\mathop{\mathrm{tr}}}

\newcommand{\fatops}[2]{\genfrac{}{}{0pt}{2}{#1}{#2}}   
\newcommand{\scl}{\fatops{ \raisebox{-.3cm}{$\textstyle\sim$} }
                         { \scriptstyle\hbar\to 0             }
                 }

\newcommand{\ket}[1]{|#1\rangle} 
\newcommand{\kete}[1]{|\kern.3ex#1\kern.3ex\rangle}

\newcommand{\brae}[1]{\langle\kern.3ex #1 \kern.3ex|} 
\begin{document}

\title{Importance of the  Wick rotation on Tunnelling }
\author{Amaury Mouchet}
\affiliation{Laboratoire de Math\'ematiques
  et de Physique Th\'eorique, Universit\'e Fran\c{c}ois Rabelais de Tours --- \textsc{\textsc{cnrs (umr 6083)}},
F\'ed\'eration Denis Poisson,
 Parc de Grandmont 37200
  Tours,  France.}

\email[email: ]{mouchet@phys.univ-tours.fr}

\date{\today}


\begin{abstract}
A continuous complex rotation of time~$t\mapsto t\EXP{-i\theta}$ is shown to 
smooth out the huge fluctuations that characterise chaotic tunnelling. 
This is illustrated in the kicked rotor model (quantum standard map) where the period
of the map is complexified:
the associated chaotic classical dynamics, if significant for~$\theta=0$, is blurred out
 long before the Wick rotation is completed ($\theta=\pi/2$). The influence of
resonances on tunnelling rates weakens exponentially as $\theta$ increases from zero, all the more rapidly
the sharper the fluctuations. The long range fluctuations can therefore be identified in a deterministic way
without ambiguity. When the last ones have been washed out, tunnelling recovers the (quasi-)integrable 
exponential behaviour governed
by the action of a regular instanton.
\end{abstract}

\maketitle

It is often admitted that the Wick rotation $t\to-\imat t$ provides
a straightforward route from a Minkowskian metric to an Euclidean metric
or from a zero temperature quantum model to a finite temperature
statistical model. In this note we will show that 
 even a small complex rotation of time, 
$t\to\EXP{-\imat\theta} t$
can affect drastically the spectral properties
of some operators that encapsulate the quantum dynamics. In particular, 
the \emph{fluctuations} of tunnelling rates and therefore some quantum transport properties,
can change by  several order of magnitude even if~$\theta\ll1$.

This work is especially motivated by the topic of chaotic tunnelling,
a theoretically challenging subject that has been resisting a
systematic and satisfying solution for more than sixteen years. Since
the pioneer quantitative
studies~\cite{Lin/Ballentine90a}, it has been widely
observed~\cite{fn:afroids} that
tunnelling, \textit{i.e.} a quantum process that is forbidden at a
classical level, exhibits a huge sensitivity to any perturbation and
generically fluctuates by several orders of magnitudes if the
underlying classical dynamics is non-integrable. Understanding,
predicting and controlling this behaviour, remain  widely open problems
for which several strategies have been proposed.  Following 
semiclassical methods, known to be successful in tackling the issues
of tunnelling in integrable systems, the most natural strategy is to
try to express a tunnelling rate (for instance a decay time or an
oscillation period between two symmetric wells) in terms of complex
solutions of Hamilton's equations only. Already highly non-trivial in the
multidimensional (non-separable) integrable or quasi-integrable
cases~\cite{Miller74a,Wilkinson86a},
this program appears even more difficult when some of the integrals of
motions are strongly broken. Indeed, it has been
discovered~\cite{Shudo/Ikeda95a} that chaos reveals
itself in the complex phase-space though some fractal structures, the
so-called Laputa islands, that look like agglomerates of complex
classical trajectories.  It is only recently that some encouraging, 
significant steps were realised in  retaining the relevant semiclassical
skeleton~\cite{Shudo+02a} for tunnelling.  To bypass this purely
semiclassical strategy, it has been
proposed~\cite{Bohigas+93a}
to replace the chaotic, though deterministic, transition amplitudes
(\textit{i.e.} involving one or more chaotic quantum states) by some
random matrix elements and hopefully compute an average tunnelling
behaviour on the appropriate statistical ensemble. This hybrid
approach, mixing together some semiclassical integrable ingredients
and statistical chaotic ones, allows to establish a precise and
quantitative connection between the quantum and classical resonances
which play a crucial r\^ole for understanding multidimensional
tunnelling~\cite{Brodier+01a}.  When classical chaos is
well developed, the overlap of classical resonances has a quantum
counterpart: the collective effect of the quasi-coincidences of
quantum frequencies, which cannot be isolated one from the other, not
only causes the huge fluctuations of a tunnelling rate but also
enhances its average
behaviour~\cite{Eltschka/Schlagheck05a,Schlagheck+06a,Mouchet+06a}.
In this context, the complex rotation of time that is proposed in this
note provides a continuous way to understand how the resonances
conspire to create a chaotic tunnelling regime.  No statistical
ingredients are required and therefore the present approach
establishes a new bridge between the two strategies described above
since it provides a deterministic,  resonances-governed transition
between a regular and a chaotic tunnelling regime where
the complex classical solutions play a natural and essential r\^ole.
 
\begin{figure}[!ht]
\center
\includegraphics[width=8.5cm]{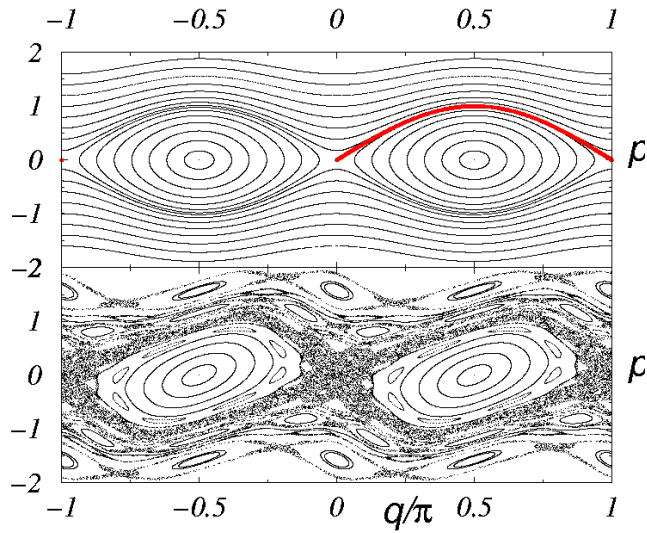}
\caption{\label{fig:poincare_kpend} Real phase space trajectories for
the kicked rotor with~$\tau$ real and~$\gamma=0.25$. The upper panel
corresponds to the quasi-integrable case ($\tau\to0^+$). The (red)
thick line is the trajectory joining the unstable fixed
point~$(p=0,q=0)$ to $(p=0,q=\pi)$ after an infinite number of
iterations of the  map.  The mixed dynamics in the lower panel
corresponds to~$\tau=1$ where a chaotic sea separates the two regular
islands in the neighbourhood of the stable points~$(p=0,q=\pm\pi/2)$. }
\end{figure}

 One of the simplest models where chaotic tunnelling is at work
corresponds to a quantum system with one degree of freedom whose
dynamics is a sequence of periodically alternating kinetic and
potential motions~\cite{Berry+79a}.  After one period~$\tau$, the
quantum evolution operator (also known as the quantum map or the
Floquet operator) is given by
\begin{equation}\label{eq:U}
  \hat{U}(\tau)\ =\ \EXP{-\imat\tau f(\hat{p})/\hbar}\EXP{-\imat\tau g(\hat{q})/\hbar}\;.
\end{equation}  The associated
 classical dynamics after one period is described by  the discrete Hamilton equations 
 (the Poincar\'e map): $(p_0,q_0)\mapsto(p_1,q_1)$ with
    $p_1=p_0-\tau g'(q_0)$
    and $q_1=q_0+\tau f'(p_1)$
(the primes stand for the derivative of the smooth functions~$f$
and~$g$).  The choice~$f(p)\DEFt p^2/2$ and $g(q)\DEFt\gamma\cos q$ where
$\gamma$ is a real parameter, corresponds to the well known kicked
rotor model (the standard map) that is extensively studied for
understanding the subtle interplay between classical and quantum
transport~\cite{Casati+79a}. It is convenient to work with the usual 
symmetric
double-well situation that can be obtained from the kicked rotor model 
by unfolding
 the cylindrical phase-space on a double spatial period and 
work in the
following with $g(q)\DEFt\gamma\cos2q$ with strictly periodic quantum
periodicity for any  $q$-translation by~$2\pi$ \cite{fn:bloch}.  The
classical dynamics is shown in
Figure~\ref{fig:poincare_kpend}.
No real trajectories connect the interior of 
the two symmetric islands  in the neighbourhood of
the two 
fixed points~$p=0$ and $q=\pm\pi/2$ that remain stable as long as
 \begin{equation}\label{eq:stability}
  \tau\ \in\ \big] -1/\sqrt{\gamma}\;,\;1/\sqrt{\gamma}\; \big[ \;.
\end{equation}
At the quantum level, we will consider the dynamics that is described in terms of the
eigenstates of~\eqref{eq:U} that are strictly invariant under a
$q$-translation by~$2\pi$. The classical phase-space
parity symmetry~$(p,q)\mapsto(-p,-q)$ allows to classify the spectrum
of~\eqref{eq:U} accordingly: we will denote~$u_n^{+}$
(resp. $u_n^{-}$) the eigenvalues corresponding to the symmetric
(resp. antisymmetric) eigenstates~$\ket{\phi^\pm_n}$, $n$ is an
quantum number labelling the doublet. For real~$\tau$,
 the quasi-energies defined modulo $2\pi\hbar/\tau$ by
$\epsilon^\pm_n\ \DEFt\ \frac{\imat\hbar}{\tau}\ln(u^\pm_n)$ are
real. For a given~$n$, when~$\ket{\phi^\pm_n}$ have their Husimi
representation localised inside the two wells, tunnelling is
characterised by the splitting
$\Delta\epsilon_n\DEFt\epsilon^-_n-\epsilon^+_n$. In the following, we
will drop the~$n$, keeping in mind that we will work with the central
doublet \textit{i.e.},  selected by the criterion of having the
maximal overlap with a coherent state that is located on the stable
fixed points~\cite{fn:exciteddoublet}. The expected semiclassical behaviour
of~$\Delta\epsilon$ in a (quasi-)integrable regime where no classical
resonance can be resolved by quantum eyes is given
by~\cite{Wilkinson86a}
\begin{equation}\label{eq:wilkinson}
   \Delta\epsilon \scl \alpha \hbar^{\nu}  \EXP{-|A|/\hbar}\;,
\end{equation}
where $\alpha$ and~$A$ are made of classical ($\hbar$-independent)
ingredients.  The exponent~$\nu$ depends on the nature (integrable or
quasi-integrable) of the dynamics.  For a 1\textsc{d} time-independent
system it is well known~\cite{Landau/Lifshitz77a} that $\nu=1$
and both $\alpha$ and $|A|$ can be interpreted in terms of
instantons~\cite{Mclaughlin72a}, \textit{i.e.}
expressed from the real trajectory joining (in an infinite time) the
two fixed points once the Wick rotation ($\theta=\pi/2$) has been
completed.  In multidimensional systems or in the case of a quantum
map, the generic presence of quantum (and classical) resonances will
drastically modify~\eqref{eq:wilkinson}. This can be understood as
follows: for~$\tau$ real, the~$u$'s accumulate on the unit circle and
two quasi-energies differing by almost an integer multiple of
$2\pi\hbar/\tau$ will make the corresponding $u$'s almost coincide and
generate small denominators in any perturbative expansion. Some
transitions between the central doublet and some excited doublets are
therefore greatly enhanced and give birth to large fluctuations when a
control parameter is varied. In figure~\ref{fig:split}, the latter is
chosen to be the inverse of the (effective) Planck constant
and~$\Delta\epsilon$ is plotted in black dots for~$\tau=1$
and~$\gamma=0.25$ corresponding to the classical dynamics in the lower
panel in figure~\ref{fig:poincare_kpend}.
\begin{figure}[!ht]
\center
\includegraphics[angle=-90,width=7cm,clip=true]{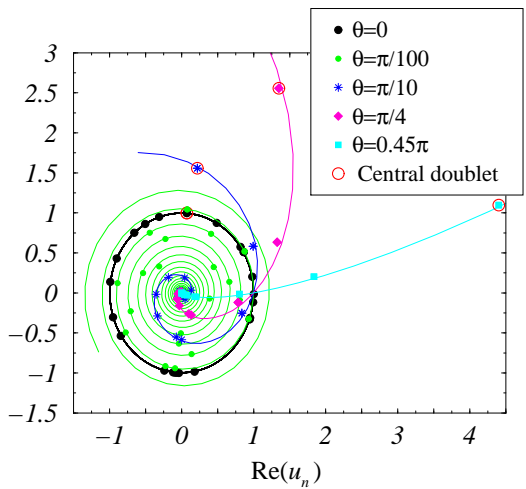}
\caption{\label{fig:spectre_spirale} (Color online) Eigenvalues~$u_n^\pm$ of~\eqref{eq:U} in the complex plane
for $\gamma=0.25$, $\hbar\simeq1/8.002$ and, $\tau=\EXP{-\imat\theta}$ with
$\theta=0$: large dots (black) ;
$\theta= 10^{-2}\pi$: small dots (green);  
  $\theta=\pi/10$: stars (blue);
 $\theta=\pi/4$: diamonds(magenta);
 $\theta=0.45\pi$: squares (cyan).
The tunnelling central doublet is circled (red). Its splitting is of order
at most~$10^{-5}$ (see figure~\ref{fig:split}) and therefore
cannot be resolved at this scale. The continuous line are the logarithmic 
spirals~$\EXP{-\imat\tau s/\hbar}$ parametrised by the real $s$. }
\end{figure} 

The operator~$\hat{U}(\tau)$ can be analytically continued in the
lower half-plane of complex~$\tau$'s. Giving a negative imaginary part
to~$\tau$ makes~$\hat{U}$ non-unitary and its eigenvalues
escape from the unit circle into the whole complex plane. In the
quasi-integrable case obtained for small enough~$|\tau|$ and~$\gamma$ for the
two exponentials in~\eqref{eq:U} to almost commute, it is expected that
the eigenvalues~$u_n^\pm$ remain near the logarithmic
spiral~$\EXP{-\imat\tau s/\hbar}$ parametrised by the real $s$. 
 It is surprisingly
the case even in a regime where classical chaos is well developed
(for $|\tau|=1$, $\gamma=0.25$, see figure~\ref{fig:poincare_kpend})
and~$\tau$ being far from the real axis.  A complex rotation of~$\tau$
by a positive angle $\theta\DEFt-\arg\tau$ appears to be a simple tool
to unfold the unit circle into a spiral and therefore to move the
resonant doublets away from each other: an increase of the quasi-energy by
$2\pi\hbar/|\tau|$ is accompanied by  a shrinking  of the modulus of
the~$u$'s by a factor of~$\exp{(-2\pi\sin\theta)}$. The larger~$\theta$,
the less the influence of the resonances on the tunnelling splitting.
This is confirmed by the plots in figure~\ref{fig:split} where we can
see how the fluctuations of~$|\Delta\epsilon|$ are progressively
smoothed by a complex time rotation. For $\theta$'s about $10^{-3}$,
only the more acute spikes are eroded. For~$\theta\simeq0.01\pi$, only
the long range fluctuations in~$1/\hbar$ have survived and the
staircase-like structure that can be observed on the semi-logarithmic
plot is very similar to the average behaviour of~$\ln|\Delta\epsilon|$
that is obtained from the resonant-assisted tunnelling hybrid method
described above (compare with \cite[Figures 1 and
2]{Eltschka/Schlagheck05a} and \cite[Figure 10]{Mouchet+06a}) but here
no averaging process is required and there is no rough transition
between staircase steps due to the semiclassical truncation of the
Hamiltonian between a regular part and the random matrix that models
the transitions involving a chaotic state.
  \begin{figure}[!ht]
\center
\includegraphics[angle=-90,width=8.5cm]{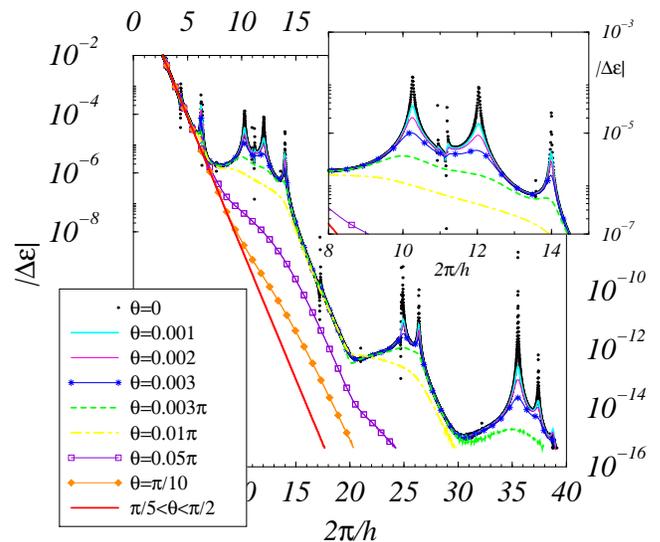}
\caption{\label{fig:split} (Color online) Modulus of the central tunnelling splitting for the kicked rotor
with $\gamma=0.25$, $\tau=\EXP{-\imat\theta}$.
$\theta=0$: dots (black) ;
 $\theta=10^{-3}$: thick solid line (cyan) ;  
$\theta=2\ 10^{-3}$:  thin solid line (magenta);
 $\theta=3\ 10^{-3}$: starred solid line (blue);
 $\theta=3\ 10^{-3}\pi$: dotted line (green);  
$\theta= 10^{-2}\pi$: dashed line (yellow);  
$\theta=5\ 10^{-2}\pi$: squared line (violet);
  $\theta=\pi/10$: diamond line (orange); 
$\pi/5\leqslant\theta\leqslant\pi/2$: thick solid line (red). }
\end{figure} 
\begin{figure}[!ht]
\center
\includegraphics[angle=-90,width=8.5cm]{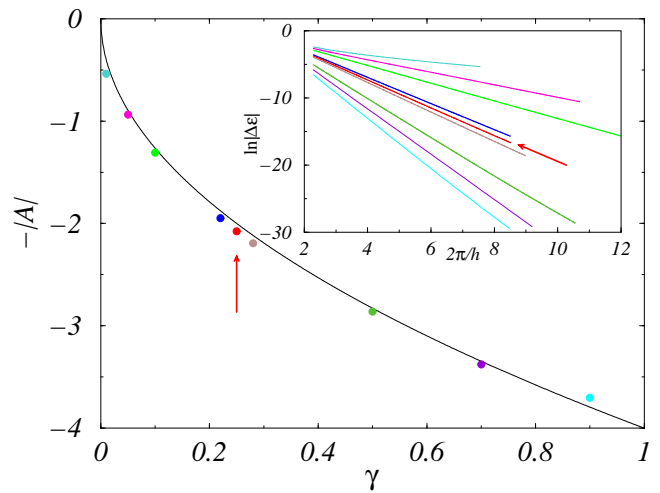}
\caption{\label{fig:penteslimites_kpend} (Color online) As~$\theta$
increases, the semi-logarithmic graphs
of~$1/\hbar\mapsto|\Delta\epsilon|$ eventually accumulate on  straight
lines (shown in the inset) whose slope~$-|A|$
(see~\eqref{eq:wilkinson}) is reported for several values of~$\gamma$
($|\tau|=1$ being fixed). The red arrows indicate the
case~$\gamma=0.25$ corresponding to figures~\ref{fig:poincare_kpend}
and~\ref{fig:split}. The continuous black line is the graph of~$-4\sqrt{\gamma}$}
\end{figure}
What is unexpected and of course requires further future
investigations, is that the splitting curves (in semi-log plots)
converge quickly, say for~$\pi/5\lesssim\theta\leq\pi/2$, to a straight
line in agreement with the (quasi-)integrable
behaviour~\eqref{eq:wilkinson}. The slope $-|A|$ is roughly the same
as the one that appears in between the plateaus for
smaller~$\theta$. Moreover, keeping $|\tau|=1$ and computing the slope
of the limiting straight lines obtained for several values
of~$\gamma$, including the ones for which the stable islands are
almost dissolved in the chaotic sea for real~$\tau$ (according to
\eqref{eq:stability} the bifurcation point where the stable fixed
points lose their stability occurs at~$\gamma=1$).  It is shown in
figure~\ref{fig:penteslimites_kpend} that~$|A|\simeq 4\sqrt{\gamma}$, 
which can be interpreted as the action of the instanton of the regular
regime.  It is easy to check that the complete Wick rotation
($\theta=\pi/2$) of the kicked rotor corresponds to a simple
$q$-translation by~$\pi/2$ of the dynamics
obtained for~$\theta=0$.  Therefore the action of the instanton
that joins~$(p=0,q=\pm\pi)$ in an infinite purely imaginary time is the
area~$4\sqrt{\gamma}$ under the separatrix shown in the upper panel of
figure~\ref{fig:poincare_kpend}.
From  a purely semiclassical point of view, as recalled above, one needs
to elucidate the relevant structures in the complex phase-space.
For~$\theta>0$, those are expected to be more tractable, compared to
the situation where~$\theta=0$. It can be predicted without too much
risk that the inextricable fractal structure of the Laputa islands
(for the kicked rotor see Figure~2 of \cite{Shudo+02a}) simplifies
 when increasing~$\theta$. In other words,
decreasing~$\theta$ towards~$0$ should allow us to understand the
formation of the delicate pattern of the Laputa chains from isolated
orbits in the same time as  we  see in figure~\ref{fig:split} how the resonances overlap and
glue together in a dense set of $\Delta\epsilon$ fluctuations.

The above analysis raises of course several (still open) questions. 
 In particular the r\^ole of the regular instanton is puzzling
since it represents a primitive classical trajectory with infinite
period and therefore seems out of reach, if one were to tentatively try to
obtain a trace formula \`a la Gutzwiller starting from the estimate
\begin{multline}\label{eq:deltatr}
  \Delta\epsilon_0\sim
\frac{2\imat\hbar}{N\tau}\frac{\tr\!\big(\hat{S}\hat{U}^N(\tau)\big)}{\tr\!\big(\hat{U}^N(\tau)\big)}
\\  =
\frac{2\imat\hbar}{N\tau}
\frac{\sum_n\left(\EXP{-\imat N\tau\epsilon_n^+/\hbar}-\EXP{-\imat N\tau\epsilon_n^-/\hbar}\right)}
     {\sum_n\left(\EXP{-\imat N\tau\epsilon_n^+/\hbar}+\EXP{-\imat N\tau\epsilon_n^-/\hbar}\right)}
\end{multline}
($\hat{S}$ is the parity operator) valid in a regime where the
integer~$N$ is sufficiently large for the central doublet to dominate
all the others but sufficiently small so
that~$N\tau\Delta\epsilon/\hbar\ll1$.  Indeed, the usual resummation techniques
 lead to an expansion arranged 
in ascending order of the length of the \emph{primitive} (pseudo-)periodic \footnote{The expansions
of the denominator and the numerator of~\eqref{eq:deltatr} combine
into pseudo-periodic orbits made from the concatenation of strictly
periodic orbits with period~$N\tau$ (denominator) with one trajectory of
length~$N\tau$ connecting two points related by parity (numerator).}. As far as tunnelling is concerned, 
I have not been able to extract any relevant information from the shortest
complex pseudo-periodic orbits.
 The chaotic character
that seems to fade away for~$0<\theta<\pi/2$,
  must also be understood better. 
In this interval, all the orbits that are stable for~$\theta=0$ or~$\pi/2$
get generically destabilised since
the traces of the monodromy matrices escape from $]-2,2[$ into the complex plane
but the mixing properties in a non-compact complex phase space may vanish as well.
The non vanishing~$\mathrm{Im}\,\tau$  has the flavour of a dissipation and the suppression
of the chaos-enhancement of tunnelling reminds of the Caldeira-Leggett model \cite{Caldeira/Leggett81a}.
   
The continuous complex rotation of time undoubtedly opens new promising
perspectives for gaining a proper understanding of chaotic tunnelling
and, as recalled in the introduction, this new approach sheds light on the subtle
non-analytical connection that may happens concerning the transport
properties of complex systems that are linked by the Wick rotation.

I thank O. Brodier, D. Delande, B. Gr\'emaud for continuous  stimulating discussions on this subject
 and S. Nicolis
for his careful reading of the manuscript.

\begin{thebibliography}{10}
\bibitem{fn:bloch}{It is of course
equivalent of retaining the $q$-antiperiodic as well as the
$q$-periodic quantum states for the standard quantum map.}
\bibitem{fn:exciteddoublet}
This is not essential : it has been checked
that the general feature described in this paper applies also to the
``first excited'' doublet.
\end{thebibliography}

\end{document}